\documentclass[letterpaper,11pt]{article}

\usepackage[latin1]{inputenc}
\usepackage[T1]{fontenc}
\usepackage{graphicx}
\usepackage{amsfonts}
\usepackage{amssymb}
\usepackage{amsmath} 
\usepackage{amsthm}
\usepackage{url}
\usepackage{rotating}

\usepackage{natbib}

\usepackage[margin=1in]{geometry}

\begin{document}

\title{SIRENE: Supervised Inference of Regulatory Networks}

\author{
Fantine Mordelet\\
Institut Curie, Paris, F-75248 France\\
INSERM, U900, Paris, F-75248 France\\
Ecole des Mines de Paris F-77300 France\\
\texttt{fantine.mordelet@ensmp.fr}
\and
Jean-Philippe Vert\\
Institut Curie, Paris, F-75248 France\\
INSERM, U900, Paris, F-75248 France\\
Ecole des Mines de Paris F-77300 France\\
\texttt{jean-philippe.vert@ensmp.fr}
}

\maketitle

\begin{abstract}
Living cells are the product of gene expression programs that involve the regulated transcription of thousands of genes. The elucidation of transcriptional regulatory networks in thus needed to understand the cell's working mechanism, and can for example be useful for the discovery of novel therapeutic targets. Although several methods have been proposed to infer gene regulatory networks from gene expression data, a recent comparison on a large-scale benchmark experiment revealed that most current methods only predict a limited number of known regulations at a reasonable precision level.
We propose SIRENE, a new method for the inference of gene regulatory networks from a compendium of expression data. The method decomposes the problem of gene regulatory network inference into a large number of local binary classification problems, that focus on separating target genes from non-targets for each TF. SIRENE is thus conceptually simple and computationally efficient. We test it on a benchmark experiment aimed at predicting regulations in \emph{E. coli}, and show that it retrieves of the order of $6$ times more known regulations than other state-of-the-art inference methods.
\end{abstract}

\section{Introduction}
Elucidating the structure of gene regulatory networks is crucial to understand how transcription factors (TF) regulate gene expression and allow an organism to regulate its metabolism and adapt itself to environmental changes. While high-throughput sequencing and other post-genomics technologies offer a wealth of information about individual genes, the experimental characterization of transcriptional cis-regulation at a genome scale remains a daunting challenge, even for well-studied model organisms. \textit{In silico} methods that attempt to reconstruct such global gene regulatory networks from prior biological knowledge and available genomic and post-genomic data therefore constitute an interesting direction towards the elucidation of these networks.

Transcriptional cis-regulation directly influences the level of mRNA transcripts of regulated genes. Not surprisingly, many \textit{in silico} methods have been proposed to reconstruct gene regulatory networks from gene expression data, produced at a fast rate by microarrays \citep{Bansal2007How}. Clustering gene expression profiles across different conditions identifies groups of genes with similar transcriptomic response, suggesting co-regulation within each group \citep{Tavazoie1999Systematic}. Clustering methods are widely used, computationally efficient, but do not easily lead to the identification of regulators for a given set of genes. Some authors nonetheless have observed that identifying similarities, or more generally mutual information between the expression profiles of a TF and of a target gene is a good indicator of regulation \citep{Butte2000Discovering,Faith2007Large-scale}. When time series of gene expression data are available, other reverse-engineering methodologies can be applied to capture the interactions governing the observed dynamics. Different mathematical formalisms have been proposed to model such dynamics, including boolean networks \citep{Akutsu2000Algorithms} or ordinary or stochastic partial differential equations \citep{Chen1999Modeling,Tegner2003Reverse,Gardner2003Inferring,Chen2005stochastic,Bernardo2005Chemogenomic,Bansal2006Inference}. Some authors have also attempted to detect causality relationships between gene expression data, be they time series or compendia of various experiments, using statistical methods such as Bayesian networks \citep{Friedman2000Using}. These methods that estimate the regulatory network by fitting a dynamic or statistical model are often computationally and data demanding.

The comparison of these different approaches and of their capacity to accurately reconstruct large-scale regulatory networks has been hampered by the difficulty to assemble a realistic set of biologically validated regulatory relationships and use it as a benchmark to assess the performance of each method. Recently, \citet{Faith2007Large-scale} compiled such a benchmark, by gathering all known transcriptional cis-regulation in \textit{Escherichia coli} and collecting a compendium of several hundreds of gene expression profiling experiments. They compared several approaches, including Bayesian networks \citep{Friedman2000Using}, ARACNe \citep{Margolin2006ARACNE}, and the context likelihood of relatedness (CLR) algorithm, a new method that extends the relevance networks class of algorithms \citep{Butte2000Discovering}. They observed that CLR outperformed all other methods in prediction accuracy, and experimentally validated some predictions. CLR can therefore be considered as state-of-the-art among methods that use compendia of gene expression data for large-scale inference of regulatory networks.

In this paper we present SIRENE (Supervised Inference of REgulatory NEtworks), a new method to infer gene regulatory networks on a genome scale from a compendium of gene expression data. SIRENE differs fundamentally from other approaches in that it requires as inputs not only gene expression data, but also a list of known regulation relationships between TF and target genes. In machine learning terminology, the method is \emph{supervised} in the sense that it uses a partial knowledge of the information we want to predict in order to guide the inference engine for the prediction of new information. The necessity to input some known regulations is not a serious restriction in many applications, as many regulations have already been characterized in model organisms, and can be inferred by homology in newly sequenced genomes. Known regulations allow us to use a natural induction principle to predict new regulations: if a gene $A$ has an expression profile similar to a gene $B$ known to be regulated by a given TF, then gene $A$ is likely to be also regulated by the TF. The fact that genes with similar expression profiles are likely to be co-regulated has been used for a long time in the construction of groups of genes by unsupervised clustering of expression profiles. The novelty in our approach is to use this principle in a supervised classification paradigm. This inference paradigm has the advantage that no particular hypothesis is made regarding the relationship between the expression data of a TF and those of regulated genes. In fact, expression data for the TF are not even needed in our approach.

Many algorithms for supervised classification can be used to transform this inference principle into a working algorithm. We use in our experiments the support vector machine (SVM) algorithm, a state-of-the-art method for supervised classification. The idea to cast the problem of gene or protein networks inference as a supervised classification problem, using known interactions as inputs, has been recently proposed and investigated for the reconstruction of protein-protein interaction (PPI) and metabolic networks \citep{Yamanishi2004Protein,Ben-Hur2005Kernel}. \citet{Bleakley2007Supervised} proposed a simple method where a local model is estimated to predict the interacting partners of each protein in the network, and all local models are then combined together to predict edges throughout the network. They showed that this method gave important improvement in accuracy compared to more elaborated methods on both the PPI and metabolic networks. Here we adapt this strategy for the reconstruction of gene regulatory networks. For each TF, we estimate a local model to discriminate, based on their expression profiles, the genes regulated by the TF from others genes. All local models are then combined to rank candidate regulatory relationships between TFs and all genes in the genome. SIRENE is conceptually simple, easy to implement, and computationally scalable to whole genomes because each local model only involves the training of a supervised classification algorithm on a few hundreds or thousands examples. 

We test SIRENE on the benchmark experiment proposed by \citet{Faith2007Large-scale}, which aims at reconstructing known regulations within \textit{E. coli} genes from a compendium of gene expression data. On this benchmark, SIRENE strongly outperforms the best results reported by \citet{Faith2007Large-scale}, with the CLR algorithm. For example, at a 60\% true positive rate (precision), CLR identifies $7.5\%$ of all known regulatory relationships (recall), while SIRENE has a recall of $44.5\%$ at the same precision level using expression profiles.

\section{System and Methods}

\subsection{SIRENE}
SIRENE is a general method to infer new regulation relationships between known TF and all genes of an organism. It requires two types of data as inputs. First, each gene in the organism needs to be characterized by some data, in our case a vector of expression values in a compendium of expression profiles. Second, a list of known regulation relationships between known TF and some genes is needed. More precisely, for each TF, we need a list of genes known to be regulated by the TF, and if possible a list of genes known not to be regulated by it. Such lists can typically be constructed from publicly available databases of experimentally characterized regulation, e.g., RegulonDB for \emph{E. coli} genes \citep{Salgado2006RegulonDB}. While such databases usually do not contain informations about the absence of regulation, we discuss in Section \ref{sec:negex} below how we generate negative examples.

When such data are available, SIRENE splits the problem of regulatory network inference into many binary classification subproblems, one subproblem being associated to each TF. More precisely, for each TF, SIRENE trains a binary classifier to discriminate between genes known to be regulated and genes known not to be regulated by the TF, based on the data that characterize the genes (e.g., expression data). The rationale behind this approach is that, although we make no hypothesis regarding the relationship between the measured expression level of a TF and its targets, we assume that if two genes are regulated by the same TF then they are likely to exhibit similar expression patterns. In our implementation, we use a SVM to solve the binary classification problems (Section \ref{seq:SVM}), but any other algorithm for supervised binary classification could in principle be used. Once trained, the model associated to a given TF is able to assign to each new gene, not used during training, a score that tends to be positive and large when it believes, based on the data that characterize the gene, that the gene is regulated by the TF. The final step is to combine all scores of the different models to rank the candidate TF-gene interactions in a unique list by decreasing score.

In summary, SIRENE decomposes the difficult problem of gene regulatory network inference into a large number of subproblems that attempt to estimate local models to characterize the genes regulated by each TF. A similar approach was proposed by \citet{Bleakley2007Supervised} to infer undirected graphs, and successfully tested on the reconstruction of metabolic and PPI networks. Here we are confronted with a slightly different problem, since the graph we wish to infer is directed and we just need to infer local models to predict genes regulated by any given TF.

\subsection{SVM}
\label{seq:SVM}
In our implementation of SIRENE, we use a SVM to train predictors for each local model associated to a TF. SVM is a popular algorithm to solve general supervised binary classification problems which is considered state-of-the-art in many applications and is available in many free and public implementations \citep{Vapnik1998Statistical,Schoelkopf2004Kernel}. The basic ingredient of a SVM is a kernel function $K(x,y)$ between any two genes $x$ and $y$, that can often be thought of as a measure of similarity between the genes. In our case, the similarity between genes is measured in terms of expression profiles. Given a set of $n$ genes $x_{1},\ldots,x_{n}$ that belong to two classes, denoted arbitrarily $-1$ and $+1$, a SVM estimates a scoring function for any new gene $x$ of the form:
$$
f(x) = \sum_{i=1}^n \alpha_{i} K(x_{i},x)\,.
$$
The weights $\alpha_{i}$ in this expression are optimized by the SVM to enforce as much as possible large positive scores for genes in the class $+1$ and large negative scores for genes in the class $-1$ in the training set. A parameter, often called $C$, allows to control the possible overfitting to the training set. The scoring function $f(x)$ can then be used to rank genes with unknown class by decreasing score, from the most likely to belong to class $+1$ to the most likely to belong to class $-1$.

The kernel $K(x,y)$ defines the similarity measure used by the SVM to build the scoring function. In our experiments we want to infer regulations from gene expression data. Each collection of gene expression data is a vector, so we simply use the common Gaussian radial basis function kernel between vectors $u$ and $v$:
$$
K(u,v) = \exp\left(-\frac{||u-v||^2}{2\sigma^2}\right)\,,
$$
where $\sigma>0$ is the bandwidth parameter of the kernel. 

Each SVM has therefore two parameters, $C$ and $\sigma$. In order to limit the risk of overfitting and positive bias in our performance evaluation that could result from an over-optimization of these parameters on the benchmark data, we simply fix them for all SVM to the unique values $C=+\infty$ and $\sigma=8$. The value $C=+\infty$ means that we train hard-margin SVM, which is always possible with a Gaussian kernel \citep{Vapnik1998Statistical}. The choice $\sigma=8$ was based on the observation that we use expression profiles for 445 microarrays scaled to zero mean and unit standard deviation, i.e., each gene is represented by a vector of dimension 445 and of length $\sqrt{445}\sim 21$. Hence the distance between two orthogonal profiles is of the order of $\sqrt{2}\times\sqrt{445} \sim 32$. We expect that a bandwidth of the order of $\sigma=8$, which puts two orthogonal profiles at about $4\sigma$ from each other, is a safe default choice. We performed preliminary experiments with different values of $C$ and $\sigma$, which did not result in any significant improvement or decrease of performance, suggesting that the behaviour of SIRENE is robust to variations in its parameters around these default values. All results below were obtained with this default parameter choice.

\subsection{Choice of negative examples}
\label{sec:negex}
SIRENE being a supervised inference algorithm, two sets of positive and negative training examples are needed for each SVM. Although regulations reported in databases such as RegulonDB can safely be taken as positive training examples, the choice of negative examples is more problematic for two reasons. First, few information is published and archived regarding the fact that a given TF is found not to regulate a given target gene. Hence there is no systematic source of negative examples for our problem. A natural choice is then to take TF-gene pairs not reported to have regulatory relationships in databases as negative examples, mixing both true negative and false negative. In that case, we are then confronted with the second problem which is that, once a hard-margin SVM is trained on positive and negative examples, it always predict significantly negative scores on negative examples used during training. As a result it is not possible to use the SVM score on genes used during training if we want to find TF-pairs that were wrongly assigned to the negative class.

To overcome this issue, we propose the following scheme. Let us suppose we want to predict whether genes are regulated of not by a given TF. All genes known to be regulated by this TF form a set of positive examples, and no prediction is needed for them. The other genes are split in 3 subsets of roughly equal size. Then, in turn, each subset is taken apart, and a SVM is trained with all positive examples and all genes in the two other subsets as negative examples. The SIRENE score for the genes in the subset left apart is the SVM prediction score on these genes, which were not used during SVM training. Repeating this loop $3$ times, we obtain the SIRENE score for all genes with no known regulation by the TF. This process is then repeated for all other TF one by one. The advantage of this procedure is that, even though there are false negative in the training set of each SVM, the predictions on the genes not used during training can still be positive if some of these genes look similar to the positive training examples. 

\subsection{CLR}
We compare the performance of SIRENE with CLR, a method for gene network reconstruction from gene expression data that was shown by \citet{Faith2007Large-scale} to be state-of-the-art on a large-scale benchmark evaluation. CLR an extension of the relevance networks class of algorithm \citep{Butte2000Discovering}, which predict regulations between TF and genes when important mutual information can be detected. In the case of CLR, an adaptive background correction step is added to the estimation of mutual information. For  each gene, the statistical likelihood of the mutual information score is computed within its network context. Then, for each TF-target gene pair, the mutual information score is compared to the context likelihood of both the TF and the target gene, and turned into a $z$-score. Putative TF-gene interactions are then ranked by decreasing $z$-score.

\subsection{Experimental protocol}
\label{sec:expprotocl}
\begin{figure}[!tpb]
\begin{center}
\includegraphics[width=3.5cm, height=5cm]{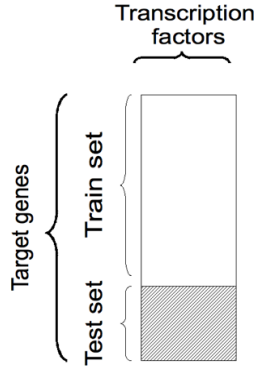}
 \caption{Cross validation for the transcriptional regulatory graph}
  \label{fig:Cvscheme2}
  \end{center}
\end{figure}

In order to assess the performance of SIRENE as an inference engine, and compare it with other existing methods, we test it on a benchmark of known regulatory network. However, SIRENE being a supervised method, we adopt a cross-validation procedure to make sure that its performance is measured on prediction not used during the model training step. Consequently we adopt the following 3-fold cross validation strategy, coherent with the SIRENE protocol to make predictions explained in Section \ref{sec:negex}. Given a set of TF, a set of genes, and a set of known TF-gene regulations within these sets, we split randomly the set of genes in $3$ parts, train the SVM for each TF on two of these subsets, and evaluate their prediction quality on the third subset, i.e., on the regulations of those genes that were not used during training (Figure~\ref{fig:Cvscheme2}). This process is repeated $3$ times, testing successively on each subset, and the prediction qualities of all folds are averaged.

In this cross-validation procedure, a particular attention must be paid to the existence of transcription units and operons in \emph{E. coli}. Indeed, a given TF typically regulates all genes within an operons, which moreover usually have very similar expression profiles. As a result, if genes within an operon are split between a training and a test set, then the SVM prediction is likely to be correct simply because the SVM will predict that a test gene with a profile very similar to a training gene should be in the same class. In other words, the SVM can probably easily recognize operons and make correct predictions due to the presence of operons. However we are interested here in the prediction of inference of regulations for new operons. To simulate this problem in our cross-validation setting, we make sure that all genes that belong to the same operon are in the same subset of genes, i.e., are always either in the training set or in the test set together. In our experiments below we report results both an a classical cross-validation setting, and on this particular scheme that preserves the integrity of operons in the train/test splits.

The CLR algorithm is evaluated with the same protocol. However, since CLR is unsupervised, the training set is not used in each fold, and the final ROC and precision/recall curves are equivalently obtained by computing the curves on all genes simultaneously.

To evaluate the quality of a prediction we rank all possible TF-gene regulation in the test set by decreasing score, and compute both the receiving operating characteristic (ROC) curve and the precision/recall (PR) curve. The ROC curve plots the recall, i.e., the percentage true interactions that have a score above a threshold, as a function of the false positive rate, i.e., the fraction of negative interactions that have a score above a threshold, when the threshold varies. The PR curve plots the precision, i.e., the percentage of true positive among the predictions above a threshold, as a function of recall, when the threshold varies. One ROC and PR curve is obtained in each fold of cross-validation, and these curves are averaged over the three folds to yield the final estimated ROC and PR curve.

\section{Data}

We used in our experiments the expression and regulation data made publicly available by \citet{Faith2007Large-scale} for \emph{E. coli}, and downloaded from {\scriptsize \url{http://gardnerlab.bu.edu/netinfer_plos_2007/?page_id=5}}. The expression data consist of a compendium of 445 \emph{E. coli} Affymetrix Antisense2 microarray expression profiles for 4345 genes. The microarrays were collected under different experimental conditions such as PH changes, growth phases, antibiotics, heat shock, different media, varying oxygen concentrations and numerous genetic perturbations. The expression data for each gene were normalized to zero mean and unit standard deviation. The regulation data consist of 3293  experimentally confirmed regulations between 154 TF and 1211 genes, extracted from the RegulonDB database \citep{Salgado2006RegulonDB}.

We downloaded the list of 899 known operons in \emph{E. coli} from RegulonDB. Each operon contains one or several genes, and each gene belongs to at most one operon. Genes not present in any of the regulonDB were considered to form an operon by themselves, resulting in a total of 3360 operons for the 4345 genes. This operon information was used to create the folds in the cross-validation procedure, as explained in Section \ref{sec:expprotocl}.

\section {Results}
\begin{figure*}[!htpb]
\begin{center}
\begin{tabular}{cc}
(A)   \includegraphics[scale=0.8]{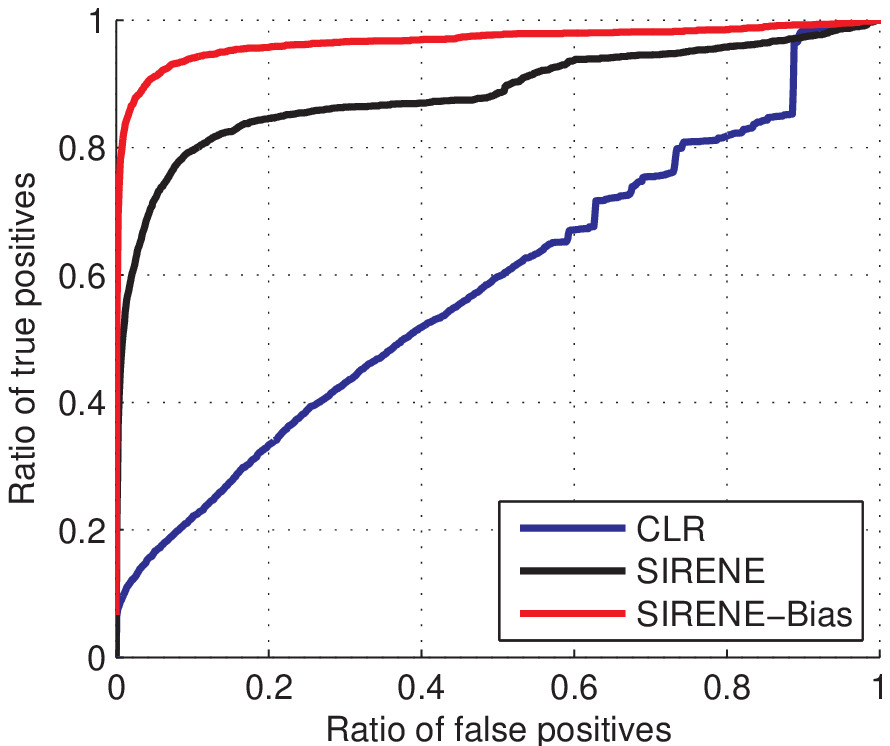} &
(B)   \includegraphics[scale=0.8]{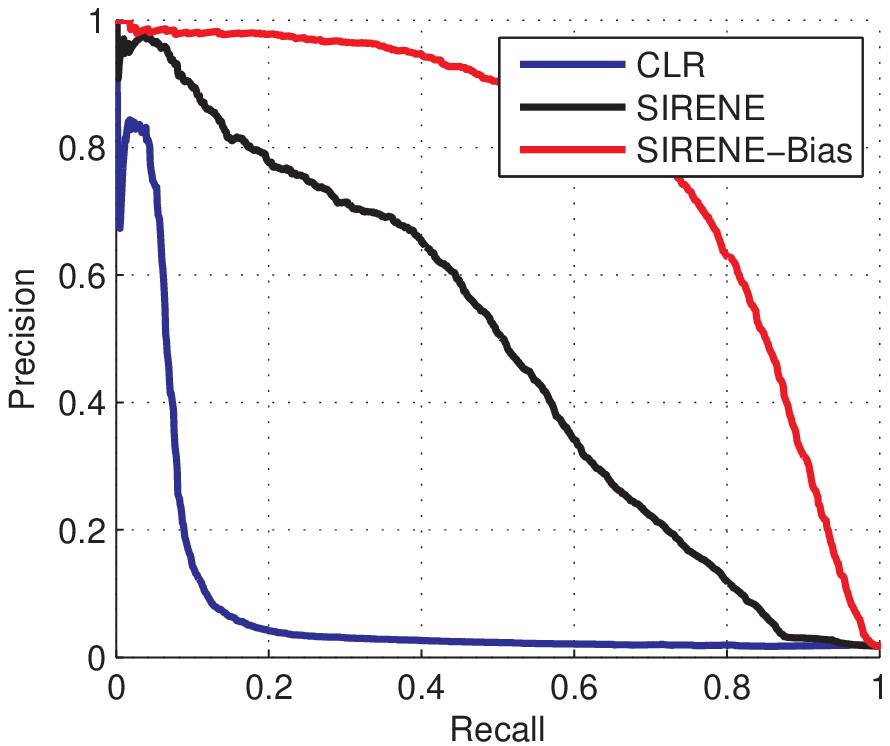} 
\end{tabular}
 \caption{Comparison of performance between CLR and SIRENE. (A) ROC curves, and (B) precision/recall curves. The SIRENE curve corresponds to the SIRENE algorithm evaluated by 3-fold cross-validation, when genes within an operon are never split between the training and the test set. The SIRENE-bias curve is the same algorithm evaluated by classical 3-fold cross-validation, where genes are randomly assigned to training and test sets.}
 \label{fig:perf}
 \end{center}
\end{figure*}

SIRENE was compared to CLR and other algorithms on the \emph{E coli} benchmark used by \citet{Faith2007Large-scale} and described in the previous section. Figure~\ref{fig:perf} shows the ROC and PR curves of CLR and SIRENE. The two curves for the later, labeled SIRENE and SIRENE-Bias, are respectively obtained when we use the cross-validation protocol presented in Section \ref{sec:expprotocl} and when we use a classical cross-validation scheme where genes within a known operon can be split between training and test sets. 

CLR scores were obtained directly from \citet{Faith2007Large-scale}. The PR curve of CLR is similar to that presented by \citet{Faith2007Large-scale}, confirming that we use the exact same benchmark. Both for ROC and PR, SIRENE performance curves are significantly above CLR. SIRENE-bias is itself much better than SIRENE, confirming the importance of the evaluation bias if operons are split artificially between training and test sets in the cross-validation procedure. In what follows we restrict ourselves to the analysis of the results of SIRENE in the correct cross-validation protocal.

The PR curve is particularly relevant because the number of true regulations is very small compared to the total number of possible TF-gene pairs. We see that the recall obtained by SIRENE, i.e., the proportion of known regulations that are correctly predicted, is several times larger than the recall of CLR at all levels of precision. More precisely, Table \ref{tab:results} compares the recalls of SIRENE, CLR and several other methods at $80\%$ and $60\%$ precision. The other methods reported are relevance network \citep{Butte2000Discovering}, ARACNe \citep{Margolin2006ARACNE}, and a Bayesian network \citep{Friedman2000Using} implemented by \citet{Faith2007Large-scale}. The performance of these three methods was taken directly from \citet{Faith2007Large-scale}.

\begin{table}[htdp!]
\caption{Recall of different gene regulation prediction algorithm at different levels of precision ($60\%$ and $80\%$). The values for relevance network, ARACNe and Bayesian network were taken from  \citet{Faith2007Large-scale}.}
\begin{center}
\begin{tabular}{|l|c|c|c|c|}
\hline
Method & Recall at 60\% & Recall at 80\%\\
\hline
SIRENE & \bf 44.5\% & \bf 17.6\% \\
\hline
CLR & 7.5\% & 5.5\% \\
\hline
Relevance networks & 4.7\% & 3.3\%  \\
\hline
ARACNe & 1\% & 0\%  \\
\hline
Bayesian network & 1\% & 0\%  \\
\hline
\end{tabular}
\end{center}
\label{tab:results}
\end{table}

At $60\%$ precision, SIRENE predicts $6$ times more known regulations than CLR, which was the best among all methods tested on this benchmark by \citet{Faith2007Large-scale}. With $44.5\%$ recall at this precision level, the performance of SIRENE allows one, in principle, to retrieve almost half of all known regulations.

\begin{figure*}[htbp!]
\begin{center}
\begin{tabular}{cc}
(A) \includegraphics[width=0.45\textwidth]{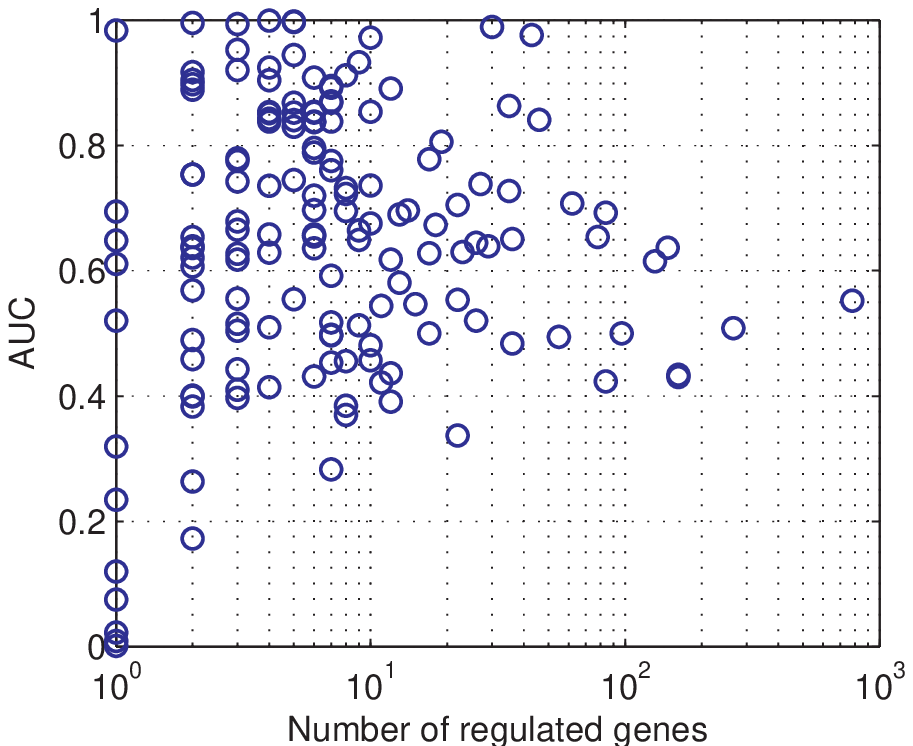}
& (B) \includegraphics[width=0.45\textwidth]{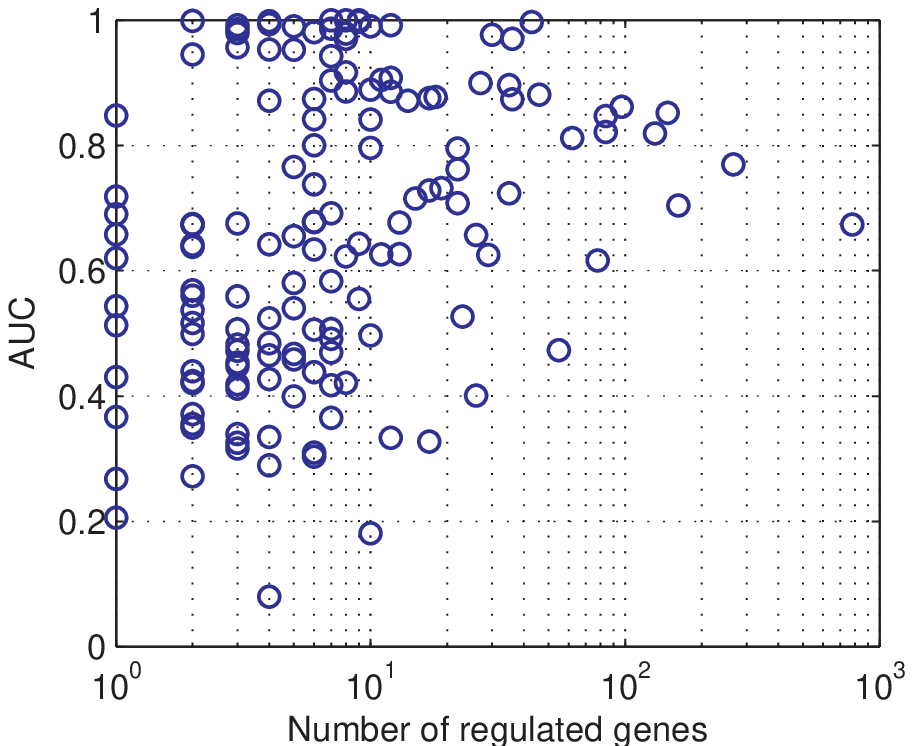}
\end{tabular}
 \caption{AUC per TF as a function of the number of regulated genes. (A) CLR and, (B) SIRENE \label{fig:auc-tf}}
\end{center}
\end{figure*}

The main conceptual difference between SIRENE and other methods is that SIRENE is a supervised method that requires known regulations to train its models. As an attempt to understand why the performance of SIRENE was better than that of other state-of-the-art unsupervised methods, we reasoned that TF with a large number of known regulated target genes could better take advantage of the supervised setting, and therefore that predictions for these TF should in general be better than predictions for TF with few known targets. To validate this hypothesis, we computed the ROC curve for SIRENE by cross-validation, restricted to the prediction of targets for each individual TF in turn. For each TF, we then computed the area under the ROC curve (AUC) as an indicator of how well the targets of each particular TF are predicted. We did this estimation for both CLR and SIRENE, and show in Figure \ref{fig:auc-tf} the distributions of AUC scores for all TF as a function of the number of known target genes in RegulonDB, for both CLR and SIRENE. As expected, the values for SIRENE tend to be larger than those for CLR. More importantly, we observe in the SIRENE plot a trend to have better AUC values for TF trained on more known targets. This trend is not present for CLR, which does not benefit from the knowledge of more or less targets for each TF. This result was expected and suggests that, as our knowledge expands and the number of known regulations continues to increase, so will the performance of supervised methods like SIRENE.

Having validated the relevance and performance of SIRENE on the regulonDB benchmark, we performed a global prediction of the \emph{E. coli} regulatory network at $60\%$ precision in order to predict new regulations in \emph{E. coli}. More precisely, for each of the 154 TF with at least one known target in RegulonDB we computed the SIRENE score for all \emph{E. coli} genes (4345 in total) that were not known targets, using the protocol described in Section \ref{sec:negex}. The RegulonDB database contained 3293 known TF-target regulations, so we assigned a score to the $4345\times 154 - 3293 = 665837$ other candidate TF-gene pairs. From the cross-validation experiment we calibrated the level of SIRENE score threshold associated to various levels of precision. We selected all pairs with a score above a threshold of $-0.41$, corresponding to an estimated precision of $60\%$. At this threshold, 991 new regulations were predicted in addition to the 3293 known ones. Combining known and predicted regulations we obtained a regulatory network with 4284 edges involving 1688 genes.

\begin{figure*}[!htpb]
\begin{center}
   \includegraphics[width=\textwidth]{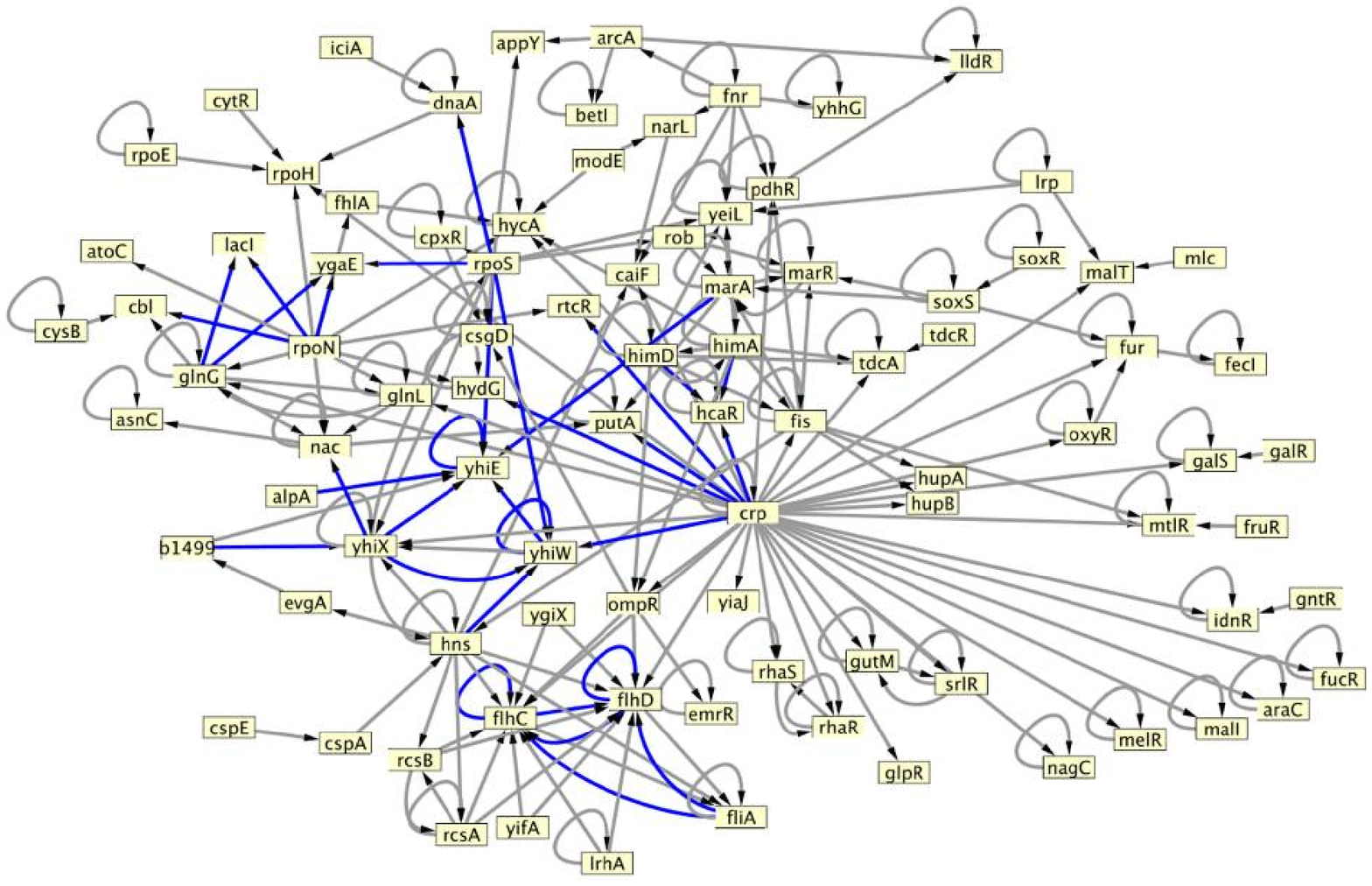}
    \caption{Main connected component of the predicted regulatory network among TF of \emph{E. coli}, at an estimated $60\%$ precision level. For clarity purpose the rpoD gene was removed from this picture. Grey arrows indicate known regulations, blue arrows indicate new predicted interactions.}
 \label{fig:prednet}
 \end{center}
\end{figure*}

In order to illustrate some predicted regulations, we focus now on the regulations of TF by other TF. Removing all non-TF genes of the predicted network, we obtain a graph with 131 TF and 349 interactions among them (TF with no interaction were removed). Among them, the \emph{rpoD} gene, which codes for the RNA polymerase sigma factor, accounts alone for 85 regulations. In order to obtain a picture easier to visualize with the Cytoscape software \citep{Shannon2003Cytoscape}, we removed rpoD from this graph, and only kept the main connected component which is shown in Figure \ref{fig:prednet}. This core regulatory network involves 90 TF, and combines 196 known regulations among them with 32 predicted ones.

Most regulations in this densely connected region of the \emph{E. coli} regulatory network have been investigated in detail, and it not a surprise that the number of newly predicted regulations is limited. Still a quick survey of the literature can confirm some of these predictions. For example, four new regulators are predicted for \emph{yhiW} (\emph{crp,hns,rpoS,yhiX} and itself), which is itself predicted to regulate \emph{yhiE}. Although these regulations were not present in the database used to train the model, they are confirmed by the literature. The GadW protein coded by \emph{yhiW} is a regulator that participates in controlling several genes of the acid resistance system. It is indeed regulated by the proteins coded by \emph{yhiX} and by the general proteins \emph{crp,hns,rpoS} that control resistance to acidity through the gad system that utilizes two isoforms of glutamate decarboxylase encoded by gene regions \emph{gadA} and \emph{gadB} and a putative glutamate:-aminobutyric acid antiporter encoded by \emph{gadC} \citep{Tucker2002Gene,Waterman2003Transcriptional,Ma2003GadE}. Another predicted regulation that was confirmed by a literature search is the dependence of \emph{hcaR}, a TF involved in the oxidative stress response, by a functional CAP protein encoded by the \emph{crp} gene \citep{Turlin2001Regulation}. Although preliminary, these first validations confirm the relevance of the approach and may suggest further experimental validations for subsystems of interest.

SIRENE is easy to implement and scales well to large-scale inference. Indeed, the main idea behind SIRENE is to decompose the network inference into a set of local binary classification problems, aimed at discriminating targets from non-targets of each TF. Although we used a SVM as a basic algorithm to solve these local problems, any algorithm for pattern recognition may be used instead. Each local problem involves at most a training set of a few thousands genes, easily manageable by most machine learning algorithms. This strategy also paves the way to the use of other genomic data to predict regulation. Indeed, local models for gene classification often improve in performance when several data, such as phylogenetic or cell subcellular localization information is available, and SVM provide a convenient framework to practically perform this data integration \citep{Lanckriet2004statistical,Bleakley2007Supervised}. Another interesting features of SIRENE is its ability to predict self-regulation, that other methods have generally difficulties to deal with.

A important limitation of SIRENE is its inability to predict targets of TF with no \emph{a priori} known target. More generally, the performance of SIRENE tends to decrease when few targets are known.  Thus, for example, it can not be used to discover new transcription factors. An interesting direction of future research is therefore to extend the predictions to TF with no known target. A possible direction may be to combine the supervised approach with other non-supervised approaches in some meaningful way.

\end{document}